# Universal thermometry of solid-liquid interfacial thermal conductance

Tao Chen, Puqing Jiang[a]

*School of Energy and Power Engineering, Huazhong University of Science and Technology, Wuhan, Hubei 430074, PR China*

[a]Authors to whom correspondence should be addressed: jpq2021@hust.edu.cn

**ABSTRACT:** Solid-liquid interfacial thermal conductance (ITC) critically influences heat transport in microfluidic, electronic, and energy systems, yet most optical thermometry techniques are limited to specific metal-liquid interfaces. In this work, we introduce a universal broadband square-pulsed thermometry method that enables simultaneous quantification of ITC across a wide range of arbitrary solid-liquid interfaces, while also providing accurate measurements of nanoscale liquid-film thickness. To validate the method, we applied it to Al-water interfaces, yielding ITC values in the range of 50-55 MW m$^{-2}$ K$^{-1}$, consistent with prior studies. The technique also reveals markedly lower ITCs for glass-water (9.9 MW m$^{-2}$ K$^{-1}$) and Si-water (5.7 MW m$^{-2}$ K$^{-1}$), and further measurements on Al-silicone oil (~10 MW m$^{-2}$ K$^{-1}$) and PMMA-silicone oil (~0.4 MW m$^{-2}$ K$^{-1}$) extend the validation to highly viscous nonpolar liquids and polymer-liquid interfaces. These results highlight the capability of the method to capture thermal transport differences across diverse solid-liquid combinations. Further comparisons with acoustic/diffuse mismatch models and molecular dynamics simulations, together with theoretical analysis, highlight the influence of vibrational mismatch, wettability, and surface condition on interfacial thermal transport. This broadly applicable technique enables rapid, quantitative characterization of solid-liquid interfacial thermal transport, with broad implications for interfacial heat transfer science and technology.

Interfacial thermal conductance (ITC) at solid-liquid boundaries critically governs heat flow in micro- and nanoscale thermal management and energy conversion devices.[1,2] In systems such as microfluidic cooling and liquid-based heat dissipation in high-power electronics, thermoelectric, and phase-change storage,[3,4] ITC directly impacts cooling efficiency, steady-state temperatures, and overall energy-transfer performance. Accurate ITC values are also indispensable for predictive models of multiphase media such as nanofluids,[5] and emulsions,[6] and soft condensed matter.[7]

*Corresponding Author: jpq2021@hust.edu.cn

Extensive studies have employed time-domain and frequency-domain thermoreflectance (TDTR and FDTR) techniques to investigate ITC at metal-liquid interfaces, revealing strong dependence of ITC on interfacial chemistry and wettability, with hydrophilic surfaces exhibiting several-fold higher conductance than hydrophobic ones.[8-13] However, these approaches are largely restricted to specific metal-liquid pairs, limiting their general applicability.

Here, we introduce a broadband square-pulsed source (SPS) thermometry method[14-17] that measures ITC across diverse solid-liquid interfaces. Operating from 1 Hz to 10 MHz, it probes liquid films confined between the transducer substrate and an arbitrary solid, enabling ITC measurements for virtually any solid-liquid pair. Each frequency sweep requires only about one minute of acquisition time, and reliable results can typically be obtained using just two to three frequencies. This combination of speed and universality allows rapid and quantitative benchmarking of interfacial heat transport, opening a broadly applicable route for the design and understanding of thermal interfaces.

The SPS thermometry method employs two continuous-wave lasers: a square-wave modulated pump periodically heats the sample, while changes in probe reflectance track the transient surface temperature. By fitting the temporal response across a broad frequency range (1 Hz-10 MHz), both ITC and liquid-film thickness can be extracted. The experimental setup and heat transfer model of SPS are detailed in Supplementary Sec. S1 and Refs.[14,18].

To generalize ITC measurements beyond metal-liquid systems, we devised three sample configurations (Fig. 1). In the conventional case [Fig. 1(a)], a bulk liquid is applied to a metal transducer on glass. Both the pump and probe beams pass through the substrate and focus at the glass-metal interface. In this configuration, the relevant unknowns are the metal-liquid ITC ($G_2$) and the liquid's thermal conductivity ($k_f$) and heat capacity ($C_f$). The glass-metal ITC ($G_1$) contributes negligibly to the overall thermal resistance, since solid-solid interfaces are typically an order of magnitude more conductive than solid-liquid ones, and is therefore treated as a fixed parameter in the analysis.

To extend ITC measurements to arbitrary solid-liquid pairs, the liquid can be confined between two substrates [Fig. 1(b)] or coupled through a coated thin film on the second substrate [Fig. 1(c)]. In these configurations, the liquid layer must be thermally thin so that heat penetrates across the film and the transient response becomes sensitive to the ITC of the top solid-liquid interface. A liquid film is considered thermally thin when its thickness is less than three times the thermal penetration depth, $d_p = \sqrt{k_f/\pi f_0 C_f}$, where $k_f$ and $C_f$ are the liquid's thermal conductivity and volumetric heat capacity, respectively, and $f_0$ is the modulation frequency. For water, this criterion corresponds to thicknesses below ~200 nm at 10 MHz or ~60 $\mu$m at 100 Hz. Conversely, one can verify that the water layer is not entirely squeezed out of the interface by contradiction: if the water were completely expelled, the experimental signals would fail to match the simulations (see Supplementary Sec. S2).

Because the laser beams do not pass through the liquid, liquid optical transparency is not required. The method only requires a suitable metal transducer with high absorption at the pump wavelength and high reflectivity and thermoreflectance coefficient at the probe wavelength, while the choice of solid and liquid is otherwise unrestricted. Together, these configurations establish a universal platform for quantifying ITC across diverse material classes, including metals, ceramics, semiconductors, and polymers, in contact with different liquids.

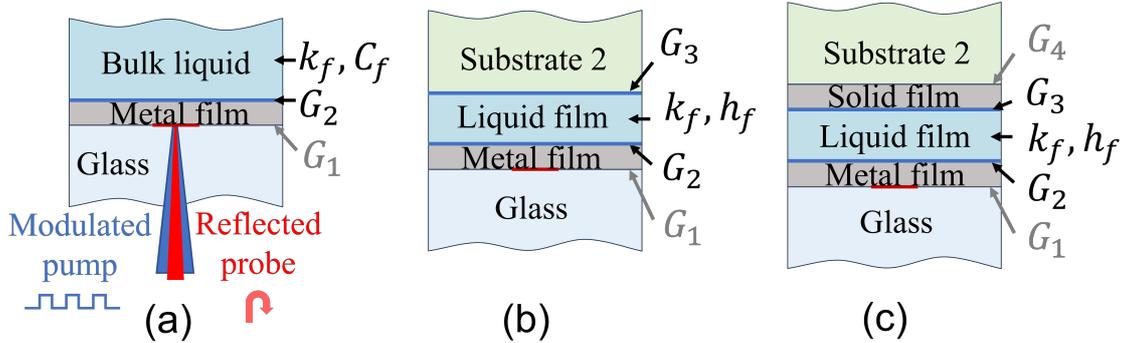

Fig. 1. Schematic diagrams of the three sample structures studied in this work: (a) a conventional setup for semi-infinite liquids, with pump and probe beams incident through a transparent glass substrate and focused at the glass/metal interface; (b) the measurement scheme for ultra-thin liquid films, where an infinitely thick substrate is used to confine the film; (c) a configuration employing a substrate coated with a metal film to confine the liquid film. In these samples, $k_f$ denotes the thermal conductivity of the liquid, $C_f$ the volumetric heat capacity of the liquid, $h_f$ the liquid thickness, and $G$ the interfacial thermal conductance, where subscripts 1, 2, 3, and 4 represent different interfaces.

To validate the method, we first measured the well-characterized Al-water interface. A calibration without liquid established the thermal properties of the glass substrate, providing reliable input parameters for subsequent ITC extraction. Applying water on Al yielded $k_f = 0.6 \pm 0.03$ W m$^{-1}$ K$^{-1}$ and $C_f = 4.1 \pm 0.2$ MJ m$^{-3}$ K$^{-1}$ for water, and $G_2 = 55 \pm 5.5$ MW m$^{-2}$ K$^{-1}$ for the Al-water interface, in excellent agreement with reported values[8,19-21] (see Supplementary Sec. S3 for details of uncertainty analysis).

Further benchmarking with a symmetric glass/Al/water/Al/glass structure [Fig. 1(c)] confirmed consistency: the extracted ITCs at both Al-water interfaces were essentially identical ($G_2 = 54 \pm 7$ MW m$^{-2}$ K$^{-1}$, and $G_3 = 50 \pm 16$ MW m$^{-2}$ K$^{-1}$), and the confined water film ($h_f = 60 \pm 8$ nm) retained bulk-like thermal conductivity ($k_f = 0.6 \pm 0.05$ W m$^{-1}$ K$^{-1}$). This agreement provides a stringent internal check and demonstrates the robustness of SPS (see Supplementary Sec. S4 for details of this benchmark study).

To demonstrate the broad applicability of our method, we measured ITC at Si-water, glass-water, and PMMA-silicone oil interfaces, where the silicone oil used had a viscosity of 100 cSt. The Si-water and glass-water cases are directly relevant to microfluidic cooling and lab-on-a-chip applications,[22,23] while the PMMA-silicone oil interface is of interest in contexts such as lubrication,[24] encapsulation,[25] and droplet-based microfluidics.[26] Details for glass-water and PMMA-silicone oil are provided in Supplementary Sec. S5, and we focus here on the Si-water case.

As shown in Fig. 2(a), the glass/Al base structure was unchanged from the benchmark case, and a polished Si wafer was used as the second substrate to confine the water layer. SPS measurements were performed at 0.5, 1, and 9 MHz, spanning thermal penetration depths from micrometer to sub-micrometer scales. These frequencies ensure that the four unknown parameters ($k_f$, $h_f$, $G_2$, and $G_3$) affect the transient thermal response in distinct ways, enabling independent extraction. The thermal conductivity and volumetric heat capacity of the Si substrate were pre-calibrated as $k_{Si} = 142$ W m$^{-1}$ K$^{-1}$ and $C_{Si} = 1.67$ MJ m$^{-3}$ K$^{-1}$ from an independent SPS experiment.

Figures 2(b)-2(g) show the experimental signals and corresponding sensitivity analyses. The sensitivity coefficient is defined as $S_\alpha = \partial(\ln R)/\partial(\ln \alpha)$, where $R$ is the measured signal and $\alpha$ the parameter of interest. A measurability analysis was performed by constructing the sensitivity matrix $\mathbf{S} = [S_{G2}, S_{kf}, S_{G3}, S_{hf}]$ and applying singular value decomposition, yielding $\Sigma$=diag(6.0, 3.0, 0.6, 0.2). Since all singular

values exceed the threshold of 0.1, this analysis confirms sufficient parameter decoupling and reliable simultaneous fitting[27,28].

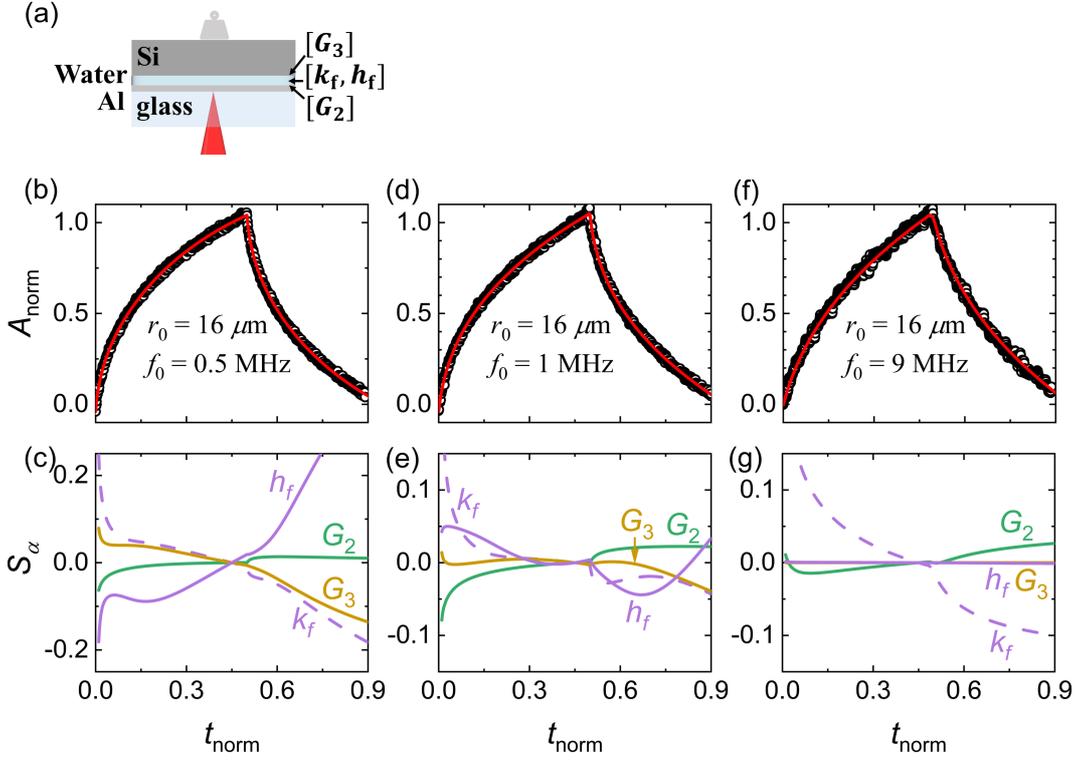

Fig. 2. Experimental configuration and results for thermal transport across the water/Si interface: (a) schematic of the sample with the water layer confined between glass/Al and Si substrates; (b, d, f) SPS signals measured at 0.5, 1, and 9 MHz, respectively, with a laser spot radius $r_0 = 16\ \mu m$; (c, e, g) corresponding sensitivity curves of the four fitting parameters at each frequency.

Analysis of sensitivity curves further supports this: at 1 MHz, $h_f$ shows a rise-fall trend while $k_f$ exhibits a rise-fall-rise pattern; the 9 MHz signal is dominated by $G_2$ and $k_f$, whereas the 0.5 MHz signal refines $G_3$. Together, these signals enable robust multi-parameter extraction.

Global fitting of the three-frequency data with a hybrid particle swarm optimization (HPSO) algorithm[29] yielded $G_2 = 52\ \text{MW m}^{-2}\ \text{K}^{-1}$, $k_f = 0.6\ \text{W m}^{-1}\ \text{K}^{-1}$, $G_3 = 5.7\ \text{MW m}^{-2}\ \text{K}^{-1}$, and $h_f = 318$ nm. The extracted $G_2$ and $k_f$ values agree with both our benchmark Al-water case and literature data,[8,21] confirming the accuracy and repeatability of SPS.

For the error analysis in this case, all measurement signals were considered simultaneously to avoid overestimating uncertainties. Specifically, the analysis incorporated high- and low-frequency signals from the glass/Al/air structure, high- and low-frequency signals from the Al/Si structure, and three sets of signals from the glass/Al/water/Si structure. Since the Al films in these samples were deposited in the same batch, their thermal conductivity, thickness, and heat capacity were assumed to be identical. The known input parameters were pre-calibrated using independent methods. The thermal conductivity of the Al film was obtained from electrical resistivity measurements by the four-point probe method[30] combined with the Wiedemann-Franz law,[31] yielding a typical uncertainty of 5%. The heat capacity of the Al film was taken from literature values with an uncertainty of 3%, and the film thickness was measured using a profilometer with an uncertainty of 2%. For water, the reported uncertainty of the heat capacity is 3%.[20] The laser spot radius was determined using the spatial-domain thermoreflectance (SDTR) method[32] with an uncertainty of 2%. The noise level of the measured signals was set to the experimentally observed value. In addition, laser power stability was verified in advance with a lock-in amplifier, and therefore, power fluctuation was not included as a source of uncertainty. Based on these settings, the resulting uncertainties in the fitted parameters were estimated as 5% for $k_{\text{glass}}$ and $C_{\text{glass}}$, 6% for $k_{\text{Si}}$ and $C_{\text{Si}}$, 14% for $G_2$, 7% for $k_f$, 6% for $h_f$, and 12% for $G_3$.

All SPS-extracted thermal properties and interface conductances for the studied configurations are summarized in Table 1, including the liquid thermal conductivity, heat capacity, film thickness, and ITCs of both the bottom Al-liquid and top solid-liquid interfaces. Consistent results between bulk- and confined-water cases validate the method, while successful measurements for chemically and mechanically dissimilar pairs (Al-oil and PMMA-oil) highlight the broad applicability of SPS to multi-interface stacks and diverse solid-liquid combinations within a unified framework.

Table 1. Summary of thermal properties of liquid and interfacial thermal conductance measured by the SPS method for different sample structures.

| Structure | $k_f$ (W m$^{-1}$ K$^{-1}$) | $C_f$ (MJ m$^{-3}$ K$^{-1}$) | $h_f$ (nm) | $G_2$ (MW m$^{-2}$ K$^{-1}$) | $G_3$ (MW m$^{-2}$ K$^{-1}$) |
|---|---|---|---|---|---|
| glass/Al/water | 0.60±0.03 (water) | 4.1±0.2 (water) | - | 55±5.5 (Al/water) | - |
| glass/Al/water/Al/glass | 0.60±0.05 (water) | - | 60±8 | 54±7.0 (Al/water) | 50±16 (Al/water) |
| glass/Al/water/Si | 0.60±0.04 (water) | - | 318±19 | 52±7.5 (Al/water) | 5.7±0.7 (Si/water) |
| glass/Al/water/glass | 0.61±0.06 (water) | - | 200±14 | 51±6.2 (Al/water) | 9.9±2.3 (glass/water) |
| glass/Al/silicone oil/PMMA | 0.11±0.01 (oil) | - | 569±36 | 10±3.8 (Al/oil) | 0.4±0.07 (PMMA/oil) |

To contextualize our results, Fig. 3(a) compares the measured ITCs with prior experimental and computational values. We obtained $G$=50-55 MW m$^{-2}$ K$^{-1}$ for Al-water, 9.9 MW m$^{-2}$ K$^{-1}$ for glass-water, and 5.7 MW m$^{-2}$ K$^{-1}$ for Si-water. Beyond aqueous systems, SPS also resolves $G$=10±3.8 MW m$^{-2}$ K$^{-1}$ for Al-silicone oil and $G$=0.40±0.07 MW m$^{-2}$ K$^{-1}$ for PMMA-silicone oil. Together, these measurements span metals, oxides, semiconductors, and polymers, and include both hydrogen-bonding and van-der-Waals liquids, underscoring the versatility of SPS in resolving ITC across distinct bonding types and phonon spectra.

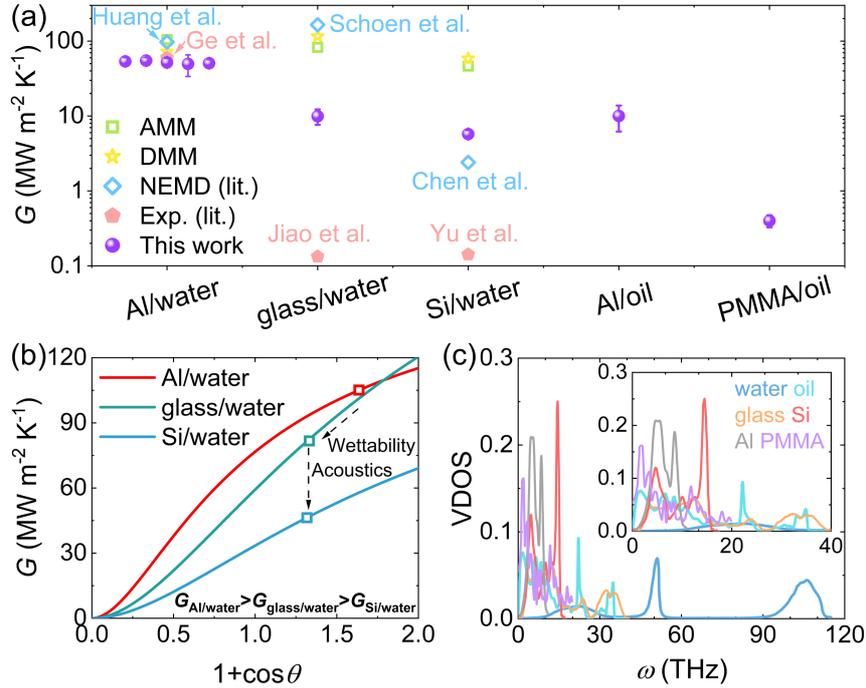

Fig. 3. (a) Measured interfacial thermal conductance ($G$) for Al-water, glass-water, Si-water, Al-silicone oil, and PMMA-silicone oil. The aqueous interfaces are benchmarked against literature data from TDTR (Ge et al.[8]), bi-directional differential 3ω (Jiao et al.[33]), and steady-state ASTM D5470 measurements (Yu et al.[34]), as well as molecular dynamics (MD) simulations of solid-water interfaces (Huang et al.[35]: Al-water; Schoen et al.[36]: Silica-water; Chen et al.[37]: Si-water). Experimental results agree well with MD data for Al-water and Si-water, while the glass-water interface shows lower conductance, likely due to surface effects. For Al-oil and PMMA-oil, no prior experimental or simulation data are available, and the absence of sound velocities precludes AMM/DMM predictions. (b) Predicted $G$ as a function of wettability (1 + cos$\theta$) based on the modified AMM model, where $\theta$ denotes the contact angle. Using the experimentally measured values of 50.5°, 70.6°, and 71.5° for the three interfaces, the model reproduces the trend Al-water > glass-water > Si-water, consistent with differences in surface wettability and acoustic mismatch. (c) Vibrational density of states (VDOS) spectra for water,[38] silicone oil,[39] glass,[40] Si,[41] Al,[42] and PMMA.[43]

Al-water results agree well with previous TDTR studies[8], while measurements on glass-water and Si-water, which are less frequently reported, reveal a strong dependence on surface conditions. For example, our glass-water conductance exceeds the values reported by Jiao et al.[33] on roughened substrates using the 3ω method, while our Si-water conductance is significantly higher than that measured by Yu and Nagayama [34] on microstructured surfaces with the ASTM D5470 method, underscoring the influence of interface quality.[44] For Al-silicone oil and PMMA-silicone oil, we did not find prior experimental or MD benchmarks for these specific pairs; accordingly, our measurements provide reference values for metal and polymer interfaces with van der Waals liquids.

Comparisons with analytical models [Fig. 3(b)] and molecular dynamics simulations show that conventional AMM and DMM generally overpredict ITC, likely because these models are based on purely particle-like assumptions and neglect coherent effects, which can lead to systematic deviations.[45] In contrast, wettability-modified models[46] capture the observed trend Al-water > glass-water > Si-water (see Supplementary Sec. S6 for details of AMM and DMM). Molecular dynamics simulations[35-37] further support these results, particularly for hydrophilic Si-water interfaces.

Beyond comparisons with analytical models and molecular dynamics, the relative magnitudes of the measured ITCs can also be rationalized by considering the vibrational spectrum overlap between the solid and liquid constituents. As shown in Fig. 3(c), the vibrational density of states (VDOS) of the materials studied in this work (from Refs. [38-43]) exhibit distinct spectral characteristics that govern the degree of interfacial coupling. Water has an exceptionally broad spectrum, spanning from sub-terahertz collective and relaxational modes to the bending (~49 THz) and O–H stretching bands (90-110 THz), which overlap strongly with the abundant low-frequency acoustic phonons of Al, yielding the highest ITC for the Al-water interface. Glass exhibits a wide phonon spectrum with a cutoff around ~20 THz and a characteristic boson peak in the terahertz range, enabling partial overlap with water's low-frequency modes and resulting in the next-highest ITC. In contrast, crystalline Si shows a main VDOS peak near 14 THz but has comparatively weak low-frequency states, and the additional presence of a native oxide and hydration layer further suppresses vibrational coupling, leading to a lower Si-water ITC. Silicone oil exhibits dominant vibrational bands in the mid-infrared (~22 THz), but also possesses low-frequency relaxational modes that can couple with Al's acoustic phonons; consequently, the Al-silicone oil ITC is comparable to that of glass-water. Finally, the PMMA-silicone oil interface yields the lowest ITC, since the primary VDOS peak of PMMA lies at ~1.7 THz and is largely offset from the dominant silicone-oil modes, producing very weak vibrational overlap and hence extremely low interfacial coupling.

Overall, the agreement among experiments, models, and theoretical analyses indicates that solid-liquid ITC is governed by vibrational overlap, wettability, and the actual interfacial structure. The consistency across metals, oxides, semiconductors, and polymers demonstrates SPS as a broadly applicable platform for benchmarking thermal transport at solid-liquid interfaces.

In summary, the square-pulsed source (SPS) method offers a robust and broadly applicable approach for quantifying ITC across diverse solid-liquid systems. Validated on the benchmark Al-water interface, it was further extended to glass-water, Si-water, Al-silicone oil, and PMMA-silicone oil pairs, revealing distinct conductance behaviors governed by surface condition, wettability, and vibrational overlap. These results demonstrate SPS as a universal platform for rapid, quantitative benchmarking of interfacial heat transport, with broad implications for thermal management, microfluidics, and energy conversion technologies.

## Supplementary Material

See the supplementary material for experimental setup, heat transfer model, uncertainty analysis, benchmark case studies, and theoretical models.

## Acknowledgment

This work was supported by the National Natural Science Foundation of China (NSFC) through Grant No. 52376058.

## AUTHOR DECLARATIONS

### Conflict of Interest

The authors have no conflicts to disclose.

## DATA AVAILABILITY

The data that support the findings of this study are available from the corresponding author upon reasonable request.